\documentclass[]{spie}  

\usepackage{comment}
\usepackage{amsmath,amsfonts,amssymb}
\usepackage{graphicx}
\usepackage[colorlinks=true, allcolors=blue]{hyperref}
\usepackage{multirow}
\usepackage{array}
\usepackage{tablefootnote}

\title{Automatic target positioning and tracking for image-guided radiotherapy without implanted fiducials}

\author[a]{Wei Zhao}
\author[a,b]{Liyue Shen}
\author[a]{Yan Wu}
\author[a]{Bin Han}
\author[a]{Yong Yang}
\author[a,b]{Lei Xing}
\affil[a]{Department of Radiation Oncology, Stanford University, Stanford, USA}
\affil[b]{Department of Electrical Engineering, Stanford University, Stanford, USA}

\authorinfo{Further author information: (Send correspondence to Lei Xing)\\Lei Xing: E-mail: lei@stanford.edu, Telephone: 1 650 498 7896 \\ }

\begin{document}
\maketitle

\begin{abstract}
Current image-guided prostate radiotherapy often relies on the use of implanted fiducials or transducers for target localization. Fiducial or transducer insertion requires an invasive procedure that adds cost and risks for bleeding, infection and discomfort to some patients. We are developing a novel markerless prostate localization strategy using a pre-trained deep learning model to interpret routine projection kV X-ray images without the need for daily cone-beam computed tomography (CBCT).
A deep learning model was first trained by using several thousand annotated projection X-ray images. The trained model is capable of identifying the location of the prostate target for a given input X-ray projection image. To assess the accuracy of the approach, three patients with prostate cancer received volumetric modulated arc therapy (VMAT) were retrospectively studied. The results obtained by using the deep learning model and the actual position of the prostate were compared quantitatively.
The deviations between the target positions obtained by the deep learning model and the corresponding annotations ranged from 1.66 mm to 2.77 mm for anterior-posterior (AP) direction, and from 1.15 mm to 2.88 mm for lateral direction. Target position provided by deep learning model for the kV images acquired using OBI is found to be consistent that derived from the fiducials.
This study demonstrates, for the first time, that highly accurate markerless prostate localization based on deep learning is achievable. The strategy provides a clinically valuable solution to daily patient positioning and real-time target tracking for image-guided radiotherapy (IGRT) and interventions.
\end{abstract}

\keywords{Image-guided radiotherapy, Image-guided intervention, Deep learning, Marker-free, Localization and tracking}

\section{INTRODUCTION}
\label{sec:intro}  

Past two decades have witnessed tremendous advances in conformal radiation therapy and new modalities such as volumetric modulated arc therapy (VMAT)  has greatly augmented our ability to shape the isodose distribution to better conform radiation dose to the tumor target while sparing the organs at risks (OARs). In practice, however, just being able to produce conformal dose distributions is not enough as it only fulfills part of the requirements of precision radiotherapy (RT).  For patients to truly benefit from the advanced planning and dose delivery techniques, we must also ensure that the planned dose is delivered to the right location and, in the case of treatment of a moving tumor, at the right time. For example, the position of prostate may change from fraction to fraction (inter-fraction prostate motion \cite{aubry2004measurements}) as well as during the dose delivery (intra-fraction prostate motion \cite{crook1995prostate,kotte2007intrafraction,adamson2010prostate}). Hence, an effective method for tumor localization, preferably in real-time, is thus of great clinical significance for the success of precision RT.

In practice, much efforts have been devoted to developing various image guidance strategies to ensure the accuracy of beam targeting \cite{xing2006overview,dawson2007advances}, which include the use of stereoscopic or monoscopic kV X-ray imaging \cite{soete2002clinical}, hybrid kV and MV imaging \cite{wiersma2008combined,liu2008real}, cone-beam computed tomography (CBCT) \cite{jaffray2002flat,barney2011image}, on-board MR imaging \cite{acharya2016online}, ultrasound imaging \cite{fontanarosa2015review}, etc. In the case of X-ray image guided RT, because of the low soft tissue contrast, it is difficult to see the tumor on the projection X-ray images. Thus, metallic fiducials are often implanted into the tumor volume or adjacent normal tissue to facilitate the patient setup and real-time tumor tracking \cite{zelefsky2012improved,campbell2017automated}. However, the implantation of the FM is an invasive procedure which introduces possible bleeding, infection and discomfort to the patient. It also needs the service of an interventional radiologist or other specialist and prolongs the treatment procedure. Besides, studies have shown FM can migrate within the patient and the prostate exhibited random deformation whose standard deviation is up to 1.5 mm, bringing uncertainty to patient setup and target localization \cite{nichol2007magnetic}.

Recently, deep learning has attracted much attention for various medical applications, such as positron emission tomography (PET)/magnetic resonance imaging (MRI) attenuation correction \cite{liu2017deep}, and segmentation of organ at risk for radiotherapy \cite{ibragimov2017,qin2018}. The purpose of this study was to investigate a novel markerless prostate localization strategy using a pre-trained deep learning model to interpret routine projection kV X-ray images for prostate IGRT.

\section{Methods and Materials}
\subsection{Deep learning for tumor target localization}


   \begin{figure} [t]
   \begin{center}
   \begin{tabular}{c} 
   \includegraphics[height=15cm]{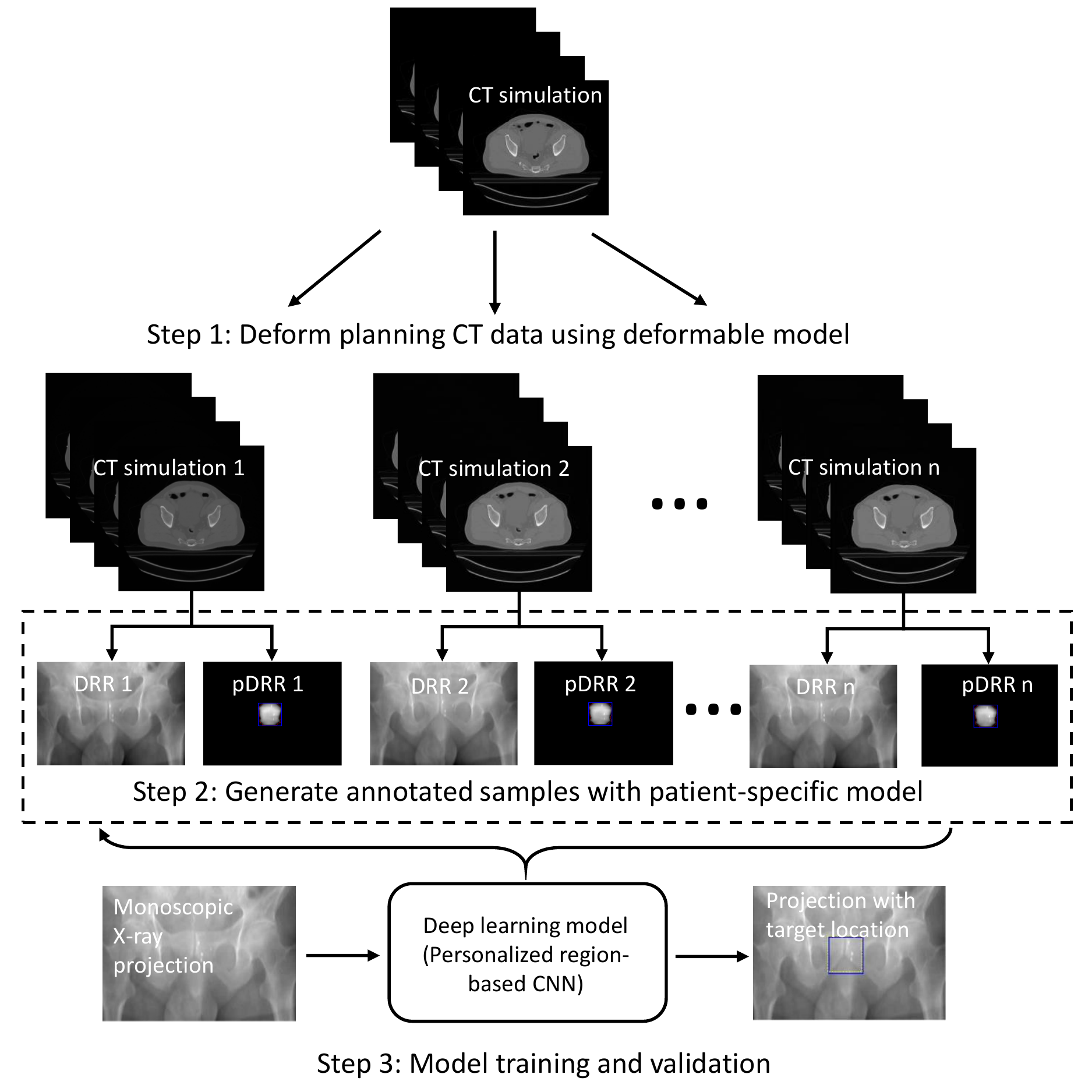}
   \end{tabular}
   \end{center}
   \caption[example]
   { \label{fig:f1}
Overall flowchart of the proposed deep learning-based treatment target localization method. Abbreviation: DRR = digitally reconstructed radiograph, pDRR = prostate-only digitally reconstructed radiograph, CNN = convolutional neural network.}
   \end{figure}

In deep learning, a computer model learns to perform prediction or classification from images or other forms of data. For the stated tumor localization problem, the following are the major tasks to accomplish in order to obtain a reliable deep learning model: construction of hierarchical neural network, collection and annotation of training datasets, training of deep learning model, and validation. Figure~\ref{fig:f1} shows the steps and workflow of the proposed deep learning-based tumor localization process. The first step is to generate training datasets of kV projection X-ray images reflecting various situations of the anatomy. For this purpose, we use robust deformable models described by motion vector fields (MVFs) to deform CT simulation to different clinical scenarios. We then generate digitally reconstructed radiographs (DRR) which can be regarded as a simulated kV projection image for each deformed CT dataset in a predefined direction using realistic imaging geometry. Finally, the annotated samples are used to train a deep learning model for subsequent localization of the prostate target. Validation tests using both independent DRR and monoscopic X-ray projection obtained from kV on-board imager (OBI) system are performed. These steps are described below in details.

\subsection{Deep learning model}

    \begin{figure} [ht]
    \begin{center}
    \begin{tabular}{c}
  \includegraphics[height=6cm]{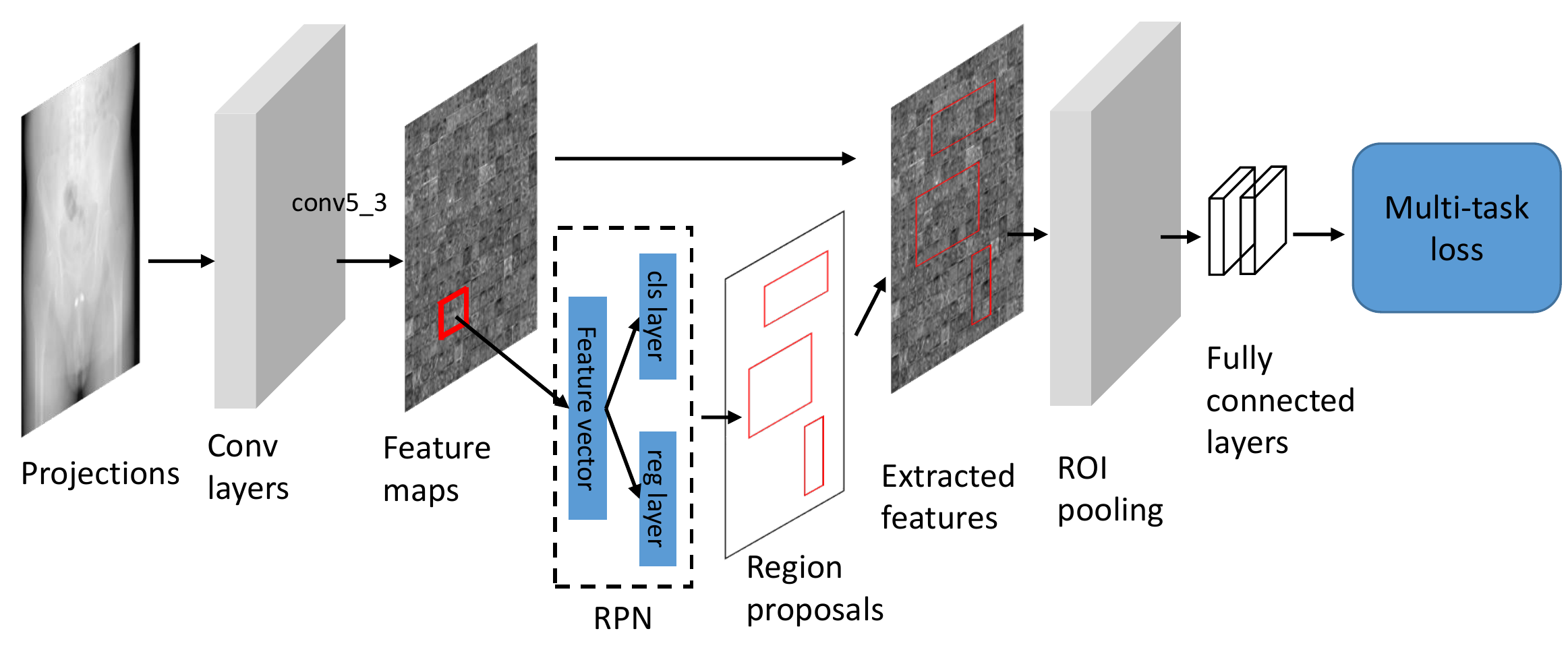}
	\end{tabular}
	\end{center}
   \caption[example]
    { \label{fig:f2}
 Schematic illustration of the deep learning model for treatment target detection. RPN = region proposed network, ROI = region of interest.}
    \end{figure}

In our deep learning model, the input is a monoscopic X-ray projection image from a given direction, say anterior-posterior (AP), or left-lateral (L-Lat), or an oblique direction obtained by the OBI system. Conventional object detection model using region-based convolutional neural network (CNN) can achieve nearly real-time detection for given region proposals \cite{girshick2015fast}. However, the calculation of the proposals is computationally intensive and presents a bottleneck for real-time detection of target region. In this study, we employ a region proposal network (RPN) to provide proposals for the region-based CNN, and the two networks share all the image convolutional features with each other \cite{ren2017faster}, as illustrated in Figure~\ref{fig:f2}. The framework jointly generates region proposal and refines its spatial location. The online kV X-ray image is first passed through CNN (13 convolutional layers) to generate its convolutional feature maps. By doing so, the CNN exploits the spatial correlation between the investigated PTV and its surrounding area and encodes it into the high-level feature hierarchies. Based on the feature maps, the RPN is construct by adding a few additional layers. Specifically, a $3\times 3$ convolutional layer map the feature hierarchies into lower-dimensional feature, which are fed into two sibling $1\times 1$ convolutional layers, a box-classification layer (cls) and a box-regression layer (reg). By providing the training DRR samples and their corresponding labels, the RPN outputs a set of region proposals, each with a score. For each of the proposals, a position sensitive region of interest pooling layer extracts a fixed length feature vector from the convolutional feature maps, which are then fed into fully connected layers for multi-task loss calculation. With this framework, the RPN and the region-based CNN share the 13 convolutional layers and ultimately enable nearly real-time accurate target detection.

\subsection{Generation of labeled training datasets for deep learning}
Training of deep learning model requires a large number of annotated datasets and this often presents a bottleneck problem in the realization of the potential of deep learning methods. Instead of collecting humongous projection kV images from the clinic, which is labor intensive and may become impractical in realistic clinical settings, we propose to generate projection kV images for different anatomical structures from deformed CT simulation for a specific patient. For each patient, CT simulation data is first deformed 3000 times to mimic patient in different statuses using MVFs, which are generated using CBCT registration. A set of 2500 deformed CT simulations is then used for model training and the remaining 500 simulations are used for model testing. The projection data for any given direction is generated by using an accurate forward X-ray projection model from the deformed CT simulations. In the forward projection calculation, we use the OBI geometry with the source to detector distance of 1500 mm and the source to patient distance of 1000 mm. In this way, the generated DRR is geometrically consistent with the kV projection image acquired using the OBI system. The DRR calculation is implemented using CUDA C with graphics processing unit (Nvidia GeForce GTX Titan X, Santa Clara, CA) acceleration. We calculate the bounding boxes of the irradiation target on DRRs and these information (top-left corner, width, height) together with the corresponding DRRs serve as annotation for deep learning model training and evaluation.

\subsection{Validation of the prostate localization model}
The trained prostate localization model is validated by retrospective analysis of 3 patients treated with VMAT. In addition to the CT simulation images, all patients received daily cone-beam CT (CBCT) scans or orthogonal kV fiducial imaging using the OBI system (Varian Medical System, Palo Alto, CA) before treatment. For each patient, we retrieved CT simulation images, target structure contours and patient setup images (CBCT images or orthogonal kV projection for different courses of treatment). For each direction of each patient, we train an independent model and the model derived prostate position is compared quantitatively with respect to the known position of the prostate. Paired-sample t-test was performed using software Matlab R2017b (Mathworks, Natick, MA) with a statistical significant level defined as p$<$0.05.

\section{Results}

   \begin{figure} [ht]
    \begin{center}
    \begin{tabular}{c}
    \includegraphics[height=12cm]{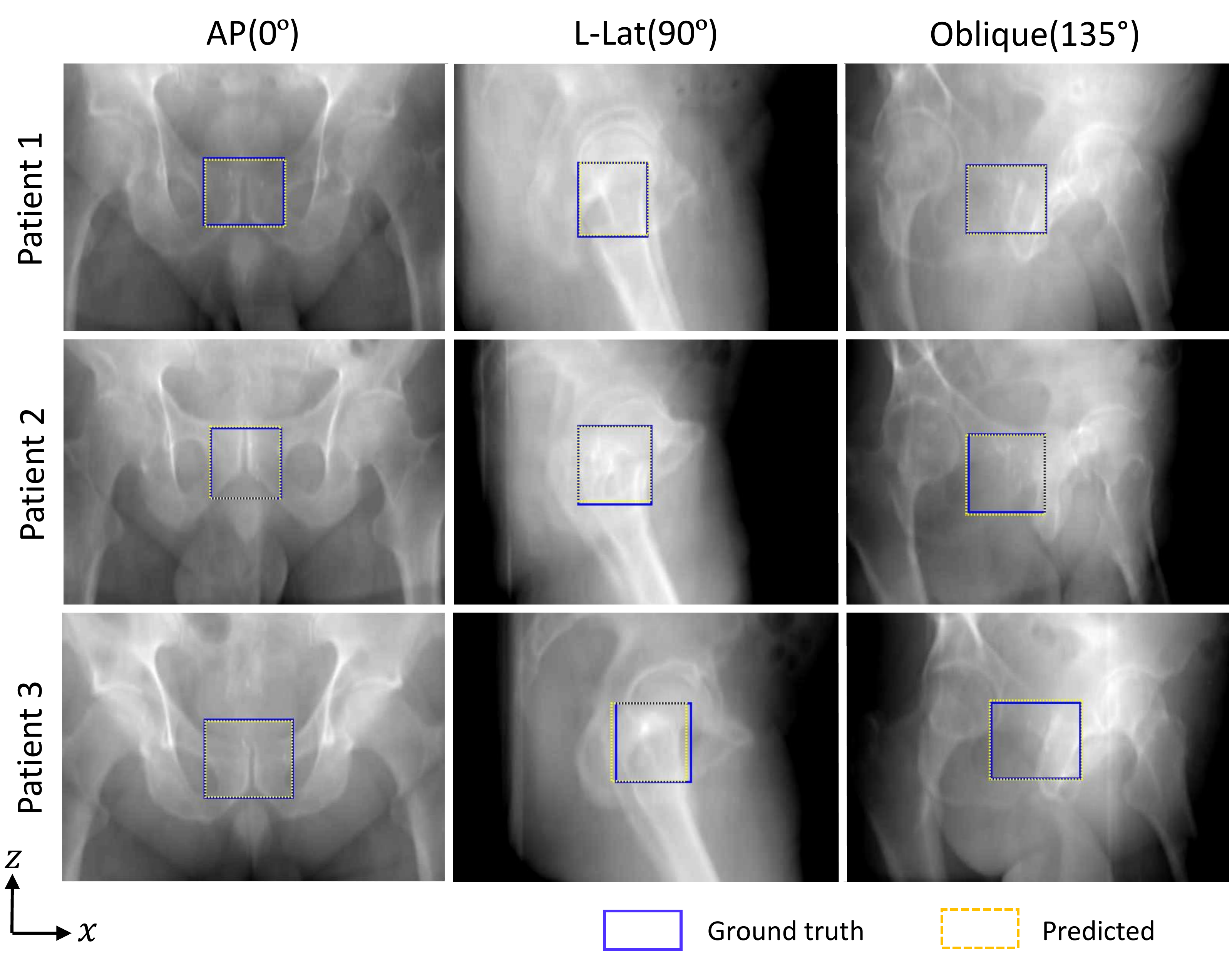}
 	\end{tabular}
 	\end{center}
    \caption[example]
    { \label{fig:f3}
 The prostate bounding boxes derived from the deep learning model (yellow dashed box) and their corresponding annotations (blue box), overlaid on top of the patient¡¯s simulated kV projection images. The first, second and third columns show the results in AP, oblique and L-Lat directions, respectively. In all directions, the predicted target position agrees with the known position better than 3mm.}
 \vspace{1em}
    \end{figure}


With patient¡¯s DRRs in three different directions as the input images, Figure~\ref{fig:f3} shows the prostate bounding boxes (dashed yellow lines) derived from the proposed approach together with the known bounding box positions (blue lines) for the three patients in AP, L-Lat, and oblique directions. Here the known positions of the prostate bounding boxes are obtained by projecting the prostate target onto the corresponding DRR planes. To facilitate visualization, the patient¡¯s DRRs after MVF deformation are presented in the figure as the background. From Figure~\ref{fig:f3}, it is seen clearly that the predicted positions match the known positions of the bounding boxes very well in all three directions. 

Quantitative comparison and statistical analyses between the deep learning model predictions and the known positions of the bounding boxes for all three patients are summarized in Table~\ref{tab:prediction 1}. For the AP and L-Lat directions which are usually used for the patient¡¯s daily setup, the agreement is better than 3 mm. Moreover, the same is true even in the oblique directions, which is important for online tracking based on the use of the kV imaging of OBI.

\begin{table}[t]
\caption[Caption for LOF]{Difference between the predicted and annotated prostate target positions in AP, L-Lat, and oblique directions.\footnotemark}
\label{tab:prediction 1}
\begin{center}
\begin{tabular}{|l|l|l|l|l|l|l|l|}
\hline
\multicolumn{2}{|c|}{\multirow{2}{*}{Patient index}} & \multicolumn{2}{ c| }{AP} &\multicolumn{2}{ c| }{LA} & \multicolumn{2}{ c |}{Oblique}\\ \cline{3-8}
\multicolumn{2}{|c|}{}
 & Deviations (mm) & P value & Deviations (mm) & P value & Deviations (mm) & P value  \\
\hline
\multirow{2}{*}{1}  & $\Delta x $  & $2.03\pm 1.47$ & 0.72 & $1.29\pm 1.48$ & 0.31 & $0.67\pm 0.87$ & 0.22\\
            & $\Delta z$  & $1.75\pm 1.90$ & 0.03 & $2.65\pm 1.54$ & 0.17 & $1.89\pm 2.43$ & 0.32 \\ \hline
 \multirow{2}{*}{2} & $\Delta x $  & $1.66\pm 1.38$ & 0.10 & $1.15\pm 0.90$ & 0.01 & $1.72\pm 0.91$ & 0.005\\
              & $\Delta z$  & $2.77\pm 2.07$ & 0.26 & $2.50\pm 2.55$ & 0.006 & $1.39\pm 0.87$ & 0.001 \\ \hline
 \multirow{2}{*}{3}  & $\Delta x $  & $1.68\pm 1.20$ & 0.96 & $1.24\pm 0.79$ & 0.64 & $1.44\pm 1.35$ & 0.13\\
              & $\Delta z$  & $2.68\pm 2.17$ & 0.04 & $2.88\pm 2.05$ & 0.001 & $2.20\pm 2.01$ & 0.43 \\ \hline
\end{tabular}
\end{center}
\end{table}



   \begin{figure} [ht]
   \begin{center}
   \begin{tabular}{c}
   \includegraphics[height=5cm]{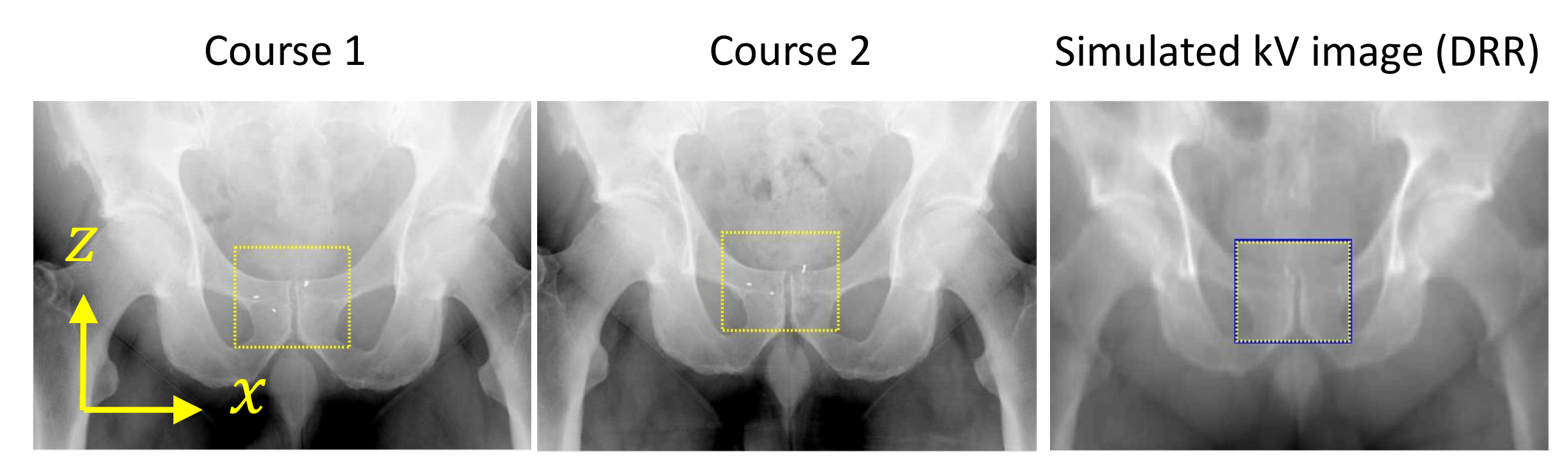}
	\end{tabular}
	\end{center}
   \caption[example]
   { \label{fig:f4}
Predicted and actual positions of the prostate target overlaid on top of the anterior-posterior simulated kV projection image (3rd column) as well as the OBI images (1st and 2nd column) for two different treatment sessions of the third patient. DRR = digitally reconstructed radiograph, OBI = on-board imager.}
   \end{figure}

Figure~\ref{fig:f4} shows the predicted target position (bounding box drawn with dashed yellow line) overlaid on top of the patient$'$s DRR and kV project images for the third case. The input kV project images are obtained using OBI before the patient$'$s VMAT treatment. In the AP direction, the deviations of the predicted position and the known position are found to be 1.68 and 2.68 mm in the x- and z-direction, respectively. This patient has three implanted fiducials and these markers afford addition assurance of the correctness of our deep learning model. In this case, the predicted prostate position is also found to be consistent with that indicated by the fiducials very well, suggesting the proposed method can provide accurate prediction of prostate position for precision RT.


\section{Discussion}
\label{sec:sections}

\footnotetext{Data are shown as means$\pm$ standard deviations. $P<0.05$ is defined as the Paired-sample t-test significance level. AP = Anterior-posterior, L-Lat = left-lateral, std = standard deviation.}

Compared with the current approaches, the proposed approach mitigates the need for implanted fiducials without compromising the target localization accuracy \cite{zhao2018redabstract,zhao2018mp,zhao2019gree}. It also enables us to localize the tumor target in the absence of visible image contrast, which represents majority of clinical situations, by effectively utilizing deep layers of image information.
Another salient feature of the proposed technique is that the training of the deep learning model does not rely on the collection of humongous data from the routine clinic practice, which has been recognized as a bottleneck in the applications of deep learning techniques \cite{hoo2016deep}. Instead, we develop a strategy of generating synthetic yet practical training datasets covering different clinical scenarios by hypothetically introducing a large number of physically realizable changes in the internal anatomy and patient positioning. This significantly simplifies the process of building a predictive model and makes it practical for a variety of clinical applications. 

\section{Conclusion}

We have proposed a deep learning model for localization of tumor target based on projection images acquired prior or during therapy. This represents the first attempt of applying deep learning to IGRT. Applications of the proposed technique to clinical prostate IGRT cases strongly suggest that highly accurate prostate tracking in projection X-ray image is readily achievable by using the proposed deep learning model.  The approach allows us to see otherwise invisible structures on the X-ray images and alleviates the need for implanted fiducials. Finally, we emphasize that the proposed method is quite broad and can be generalized to improve image guidance in many other disease sites \cite{zhao2019red}, such as the pancreas, lung, liver, brain, spine, and head and neck IGRT.




\bibliographystyle{spiebib} 

\end{document}